# Optimization of singly-charged particles identification with the AMS02 RICH detector by a machine learning method

G. Vasilev,[a,1] G. Vankova-Kirilova[a] and G. Bozhkova[a]

[a] *Sofia University St. Kliment Ohridski*
*1164 Sofia, 5 James Bourchier Blvd., Faculty of Physics, Bulgaria*
*E-mail*: georgi.danev.vasilev@cern.ch

ABSTRACT: AMS-02 is a detector currently in operation onboard the International Space Station (ISS). One of the main scientific goals of the spectrometer is the measurement of charged particle fluxes. The detector design makes possible the identification of particles and antiparticles by precise measurement of particle momentum in the AMS-02 Silicon Tracker, and velocity in the Cherenkov (RICH) detector. The RICH is able to measure the isotopic composition of the light elements (up to charge Z=5) in the kinetic energy range from a few GeV/n to about 10 GeV/n. However, the velocity reconstruction for charge 1 particles is particularly challenging due to the low number of photons they produce in the RICH detector which can lead to wrong event reconstruction. In this paper, we show the high potential of the Multilayer Perceptron deep learning model (MLP-BFGS) for identification of signal and the background due to interactions inside the AMS-02 detector, and to significantly improve particle identification by its mass.

KEYWORDS: Particle identification methods; Large detector systems for particle and astroparticle physics; Cherenkov detectors; Spectrometers;

# Contents

## 1. Introduction

The scientific interest in the light isotopes of hydrogen is determined by the question of their origin. It is believed that protons and $^4He$ nuclei are predominantly of primary origin and thus arise directly from their sources, whereas $^2H$ and $^3He$ have secondary origin, i.e. they are produced from the interactions of primary elements with the interstellar gas. Studying the deuteron flux and the corresponding secondary-to-primary ratios is essential for the understanding of propagation processes in our Galaxy and the properties of the ISM itself [1][2].

Besides the light isotopes mentioned, the detection of their corresponding antiparticles is promising as an indirect search for dark matter. Antideuterons are considered a golden sample for this search because in the energy range in which they are expected there is no background from secondary production [3].

Several experiments have been used to study singly-charged cosmic rays – IMAX [4], BESS [5], CAPRICE [6] in Earth's atmosphere, and more recently, PAMELA [7] in Earth's orbit, as well as others. Methods applied to perform isotopic identification combine the rigidity and velocity measurements, derive the mass difference between the isotopes and then separate them by using cut-based methods or fitting procedures on different variables. Particle mass is obtained from the relation $m = RZ/\beta\gamma$, with resolution

$$\Delta m/m = \sqrt{(\Delta R/R)^2 + \gamma^4(\Delta\beta/\beta)^2}$$

The mass resolution's dependence on $\gamma^4$ shows that particle identification by mass at higher energies requires very high velocity resolution. The Alpha Magnetic Spectrometer is able to provide such precise velocity measurements in the GeV-TeV energy range.

## 2. The Alpha-Magnetic Spectrometer (AMS-02)

AMS-02 is a particle physics detector currently installed and in operation on the International Space Station (ISS). Its objectives are to study the origin of antimatter, dark matter, and cosmic rays, as well as to explore new physical phenomena.

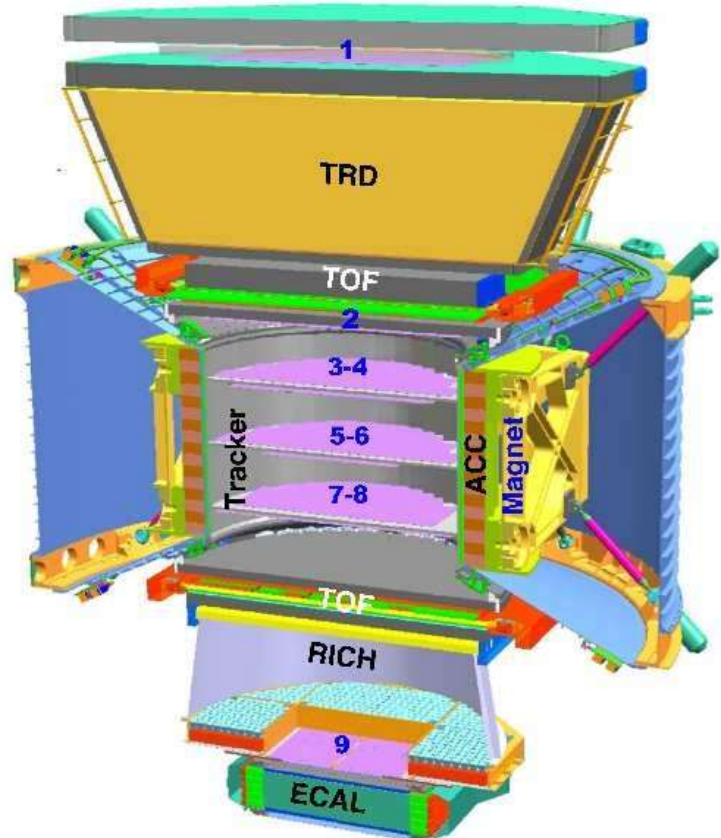

**Figure 1.** A view of the AMS-02 detector

The detector's main components [8], visualized on Figure 1, are a permanent magnet and several subdetectors which measure particle velocity ($\beta$), charge ($Z$), momentum ($p$) and rigidity ($R = p/Z$). Apart from the permanent magnet AMS-02 is composed of several subsystems: a silicon tracker consisting of nine layers, a Transition Radiation Detector (TRD), a Time of Flight (TOF) detector consisting of two pairs of scintillators above (upper TOF) and below (lower TOF) the magnet bore, a Ring Imaging Cherenkov Detector (RICH), Anti-Coincidence counters (ACC) and an Electromagnetic Calorimeter (ECAL). The permanent magnet is made up of 64 Nd-Fe-B sectors arranged in a cylinder and provides a magnetic field of 1.4 kG in its center. It is used to bend the particle trajectories as they pass through the detector.

### 2.1 The AMS-02 Subdetectors

The tracker layers measure the coordinates $(x, y)$ of particles as they pass through each layer, and these measurements serve as the basis for the reconstruction of the particle track. It also measures $Z$ through the energy deposition in the tracker layers. It consists of 9 layers – L1 through L9 – and is composed of 192 ladders containing double-sided silicon strip detectors. L2-L8 is the inner tracker. The coordinates in the tracker are measured independently with spatial resolution of $13 - 20\mu m$ in the non-bending direction (along the $x$ axis, which is directed along the main component of the magnetic field), and $5 - 10\mu m$ in the bending direction (the $y$ axis). They are used to reconstruct [9] the particle's most likely trajectory through the detector – the track. Given the track and the magnetic field map in the permanent magnet, the measurement directly yields the rigidity $R$ with resolution $\Delta R/R = 0.1$ at low rigidity (under 20 GV). Additionally, charge is measured independently in every tracker layer, with resolution depending on the type of nucleus, and then combined to yield the reconstructed charge. The charge confusion (reconstructing particles with negative charge as positive, and vice versa) in the tracker is below 8% in the energy range up to 1 TeV.

The Anti-Coincidence counters (ACCs) are 16 curved scintillator panels that surround the tracker layers located on the inside of the magnet, and they are used to filter out particles that enter the detector from the side.

The Time-of-Flight counters (TOF) precisely determines the velocity as well as the direction of the incoming particles, and also further provides a charge measurement using energy deposition in the TOF layers. Each TOF layer consists of 8 or 10 scintillating paddles. The coincidence of signals from all 4 TOF layers provides the trigger for the particle measurement. The TOF counters' average time resolution is around 160ps for $Z = 1$ particles, which corresponds to a velocity resolution $\Delta(1/\beta) \approx 0.04$. It also distinguishes upgoing from downgoing particles with a confusion probability of around $10^{-9}$.

The Transition Radiation Detector (TRD) is located above the upper TOF layers and it is used to distinguish electrons or positrons from antiprotons and protons and heavier particles, using the transition radiation produced as they pass through the detector – the lighter particles will produce substantially more transition radiation. It further determines particle charge through the measured energy loss $dE/dx$. The TRD consists of 20 layers of proportional counter tubes separated by fiber fleece radiators. The bottom and top four layers are oriented perpendicular to the middle twelve layers. The TRD separates electrons and positrons from protons. To do this, signals from the 20 layers are combined into a TRD likelihood estimator [10], which assigns values around 0.4 to positrons and electrons, and around 1.0 for protons. This allows the proton background to be rejected at a level of $10^3$ at 90% electron/positron efficiency.

The Ring-Imaging Cherenkov Detector (RICH) is located below the lower TOF layers and is used to precisely determine particle velocity and charge using the photon rings produced by the Cherenkov effect. It consists of a radiator plane made up of two radiators – a central sodium fluoride (NaF) radiator and a surrounding aerogel (Agl) radiator – which are used to detect particles in different velocity ranges. Beneath the radiators is the detector's expansion volume, and at the bottom – a photodetecting plane. This detector will be described in more detail in Section 3.

The Electromagnetic Calorimeter (ECAL) is located at the very bottom of the detector, below the RICH detector and L9. Its purpose is to optimize electron-hadron separation. It is made of a composite lead and scintillating fiber material, alternating in orientation – each two consecutive layers are oriented perpendicularly. As a particle passes through, electromagnetic showers are produced upon interaction with the lead layers over 17 radiation lengths. The scintillators then measure the shower in three dimensions in order to determine particle energy and direction. The ECAL reconstruction describes a shower using several parameters – the shower energy $E_0$, the coordinates of the shower maximum in the detector's coordinate system $X_0, Y_0, Z_0$, the coordinate $T_0$ of the shower maximum along the shower axis, and the angles $K_x$ and $K_Y$ which, together with the shower maximum, define the shower axis. These parameters are also used to separate electrons and positrons from protons using an ECAL boosted decision tree (BDT) [11][12], which assigns a value of around 1.0 for positrons and electron events, and around -1.0 for proton events. This method allows for a proton rejection efficiency of over $10^3$ in the energy range up to 1 TeV.

**2.2 Background sources**

There are several main sources of background for the AMS subdetectors. Electrons and protons are very abundant. High-energy protons can mimic electron or positron measurements in the TRD and ECAL. Charge-confused electrons and protons can contaminate positron and antiproton measurements. Backscattered and secondary particles from interactions with the detector materials can lead to wrong velocity reconstruction in TOF. Secondary particles can cause hadronic showers resembling electromagnetic ones to appear in ECAL. Electron-positron pairs from photon conversion can be detected by the TRD and ECAL. Fragmentation of heavy nuclei can cause charge misidentification. All of these effects also produce additional hits in the tracker and affect the quality of the track reconstruction. In RICH, secondary particles from upstream interactions, nuclear fragmentation and delta-rays can cause excess photons to reach the photodetecting plane, contaminating the Cherenkov ring. Therefore, accurate separation of different particle types requires effective filtering of background signals.

To reject background signals the AMS collaboration uses cut-based methods and fitting procedures supplemented with other means of distinguishing between particles, such as statistical methods like the TRD likelihood estimator [13] and machine learning methods such as the ECAL Boosted Decision Tree (BDT) [14].

## 3. Mass identification with the RICH detector

The proximity-focusing RICH detector (Figure 2) is used to determine particle velocity $\beta$ and charge $Z$. It is located in the lower half of the AMS-02, underneath the two lower TOF planes.

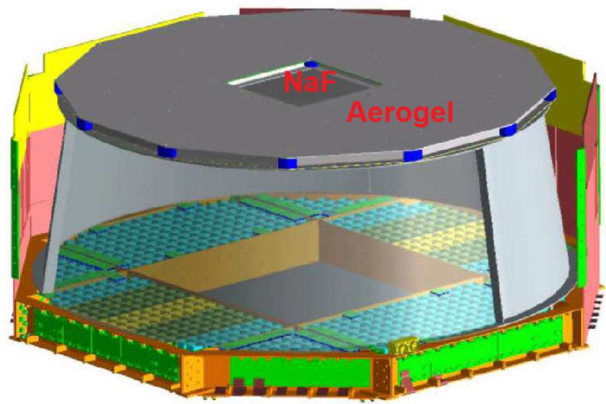

**Figure 2.** A view of the RICH detector

Its main components are two radiators, an expansion volume, and a photodetecting plane. Its design allows velocity measurements with resolution of 0.1%. The two radiators are used to measure particle velocities in different velocity ranges. The first of the two – the central radiator – is made up 16 sodium fluoride (NaF) tiles arranged in a square pattern, with refractive index $n = 1.33$. It allows for the detection of particles with velocities exceeding $\beta > 0.75$. The other radiator is made up of 92 tiles of silica aerogel surrounding the NaF radiator with refractive index $n = 1.05$. The aerogel radiator makes up around 90% of the total RICH acceptance, and it measures velocities exceeding $\beta > 0.95$. Underneath the radiators there is a large expansion volume surrounded with a high-reflectivity mirror, and on the bottom there is an arrangement of 680 multi-anode photomultiplier tubes (PMTs) [15].

When charged particles pass through the detector, Cherenkov radiation is emitted in a cone with intensity proportional to $Z^2$. The Cherenkov angle depends on the particle velocity in accordance with the relation $\theta_C = arccos(1/\beta_n)$. The data produced from the photons reaching the PMT plane, together with the particle track reconstruction from the tracker, allows for the reconstruction of the original Cherenkov angle, and from there, the particle velocity. Furthermore, the particle charge is derived from the total collected photon signal. Finally, velocity data from RICH, together with the rigidity measurement obtained from the tracker, allows for particle mass to be accurately determined.

Since the number of Cherenkov photons is proportional to $Z^2$, particles with $Z \geq 2$ have more intense, and easier to reconstruct Cherenkov rings. Conversely, for singly-charged particles the rings are very faint. The number of photon hits in a ring depends on the impact point on the radiator and only a small fraction of events produce a fully contained ring. The number of Cherenkov photons is around 5 for the NaF and around 6 for the aerogel radiator.

To reconstruct a Cherenkov ring at least three hits are required and it is obvious that if background hits near the ring are present, or if particle direction was misidentified, the velocity reconstruction will be affected. In addition, the secondary particle interactions such as delta ray production, occurring in the region between the lower tracker and the RICH can change the particle direction and introduce bias in the detected ring hits with respect to the reconstructed track from the tracker. Moreover, the aerogel is a source of photon scattering, which leads to a significant fraction of ring-uncorrelated hits. All of this causes wrong reconstruction. One can see the described effects in Figure 3, where the mass distribution in RICH for simulated proton events is shown after applying standard selection cuts.

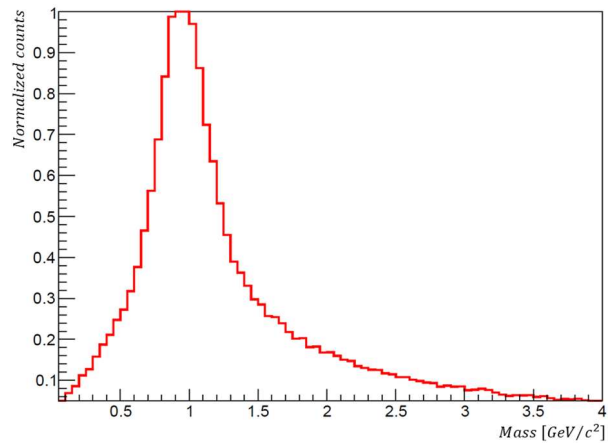

**Figure 3.** Normalized mass distribution in RICH for simulated proton events

The long tail in the distribution is due to wrong event reconstruction caused by the aforementioned effects. This is a primary source of background when attempting to make a deuteron selection.

Standard cut-based approaches can often lead to misidentification of background events as signal events, and potentially even cut real events, which is a problem when searching for rare cosmic rays. Studies on the effectiveness of Boosted Decision Tree (BDT) methods for background rejection in RICH show that over 90% of background signals can be rejected across the full velocity range in both the NaF and the aerogel radiator, with efficiency around 70% for

– 5 –

NaF and around 60% for aerogel [16]. With that in mind we search for a method capable of high background rejection with signal efficiency above 80%. The Multilayer Perceptron has shown the potential to meet such demands.

## 4. The Multilayer Perceptron

An Artificial Neural Network (ANN) [17] is a machine learning model inspired by the biological neural networks found in brains. It consists of nodes named "artificial neurons", corresponding to the brain's neurons, which are connected by "edges", corresponding to the synapses in the brain. When an external signal is introduced to the network via some input neurons, it is placed in a defined state which can then be measured from the response of the output neurons.

A neural network consisting of $n$ neurons can theoretically have $n^2$ edges. However, this means that increasing the number of neurons very quickly increases the complexity of the system. A way to reduce complexity is by organizing the network's neurons into different layers and only allowing neuron connections from one layer to the next (as shown in Figure 4) – in other words, forming a feed-forward neural network called a multilayer perceptron.

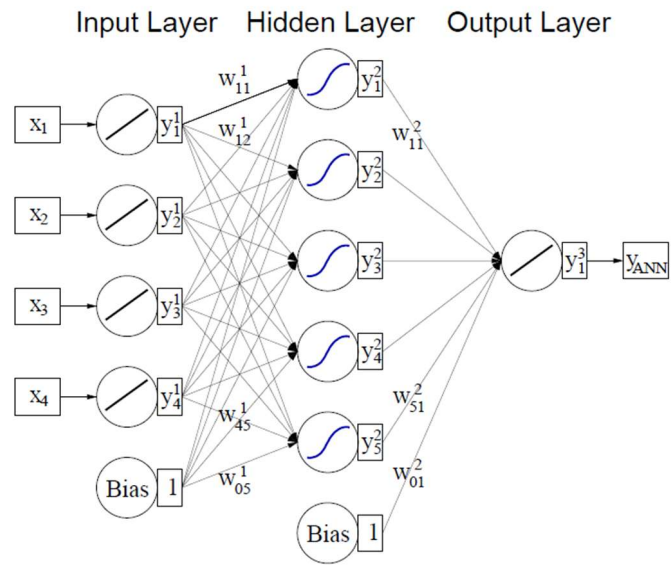

**Figure 4.** An example diagram of a multilayer perceptron network with one hidden layer

Its first layer is the input layer, the output layer is the final one, and all layers in between are hidden layers. If the network is tasked to solve a classification problem, the input layer consists of several neurons that contain the input variables, and the output layer consists of a single neuron which contains an output variable – a neural network estimator.

The Multilayer Perceptron is widely used in high energy particle physics for tasks such as classification, regression and pattern recognition. It is particularly useful in cases where particle identification is required, as is the case in this study. One key advantage of this model is its ability to model non-linear dependencies between input and output variables, which makes it a good option when faced with complex datasets where traditional analytical methods fall short.

**4.1 Training methodology and results**

For this study we use the Tools for Multivariate Analysis (TMVA) software package, available as part of the CERN ROOT package [18]. In order to train the MLP we have studied a large set of event variables from the AMS-02 detectors. From them we have selected those which are particularly sensitive to the background sources and could therefore help clean up the sample – mainly from the tracker and TOF, as well as several from RICH. We use the velocity measured in the tracker, quantities characterizing the particle's track, temporal characteristics of the particles in TOF, RICH charge, Kolmogorov probability [19] and others. The training procedure has been performed using the Broyden-Fletcher-Goldfarb-Shannon (BFGS) method [20][21][22][23] which, using the current set of variables, requires less iterations than the back-propagation method [24] and is therefore more efficient.

The samples used to train the classifier are simulation data produced by the AMS collaboration using dedicated software based on the GEANT4 package [25]. The software simulates the interactions of particles inside AMS-02, the detector response and detailed event reconstruction. To make sure that the particle velocity reconstruction in TOF is correct, only those simulated events that have recorded hits in all four scintillator planes are considered. The MLP is trained on samples of simulated proton and deuteron events. Taking into account the mass resolution of the RICH detector for signal-like events, we assume that in the case of protons, signal-like events are those with reconstructed mass within $2\sigma$ of the proton mass – $0.75 < m < 1.12 \; GeV/c^2$. Any events that do not fulfil those requirements are considered background to the signal.

Figure 5 shows as an example two of the variables used for training the MLP as mentioned in Section 4.1 – the particle velocity measured in the tracker, and the Kolmogorov probability for the NaF radiator. Signal events are shown in blue, while the background is in red.

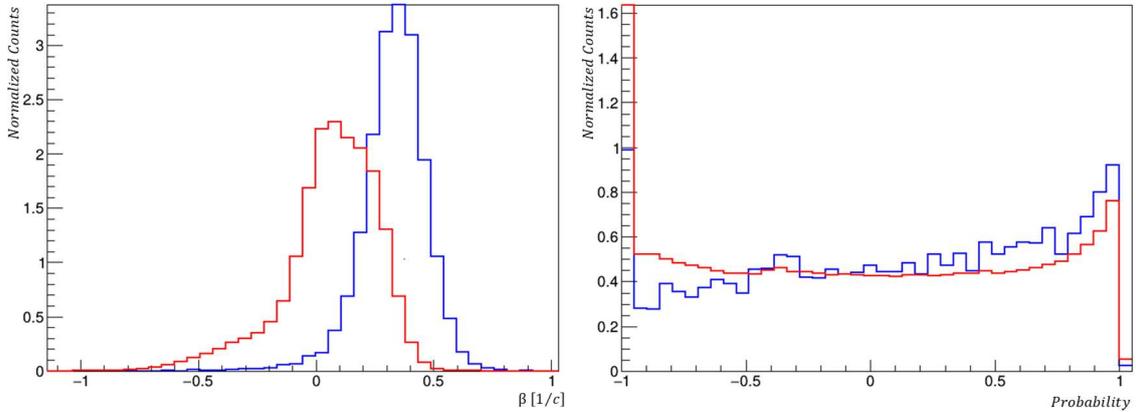

**Figure 5.** Example of two of the chosen variables for MLP training: Particle velocity measured in the tracker (left) and Kolmogorov probability (right)

The Receiver Operating Characteristic (ROC) curves for both radiators after the application of the MLP classification method on the simulated samples are shown on Figure 6. The MLP classifier shows high background rejection capabilities while simultaneously achieving 80-85% signal efficiency in both the NaF and the aerogel radiators. The ROC curves allow for the cross section significance $S(S + B)^{-1/2}$ to be easily determined. Its maximum value is the best one to use when making a cut on the MLP output. As seen on Figure 7, this is around 0.8, therefore events with output greater than 0.8 are considered signal, and anything below that value is considered background.

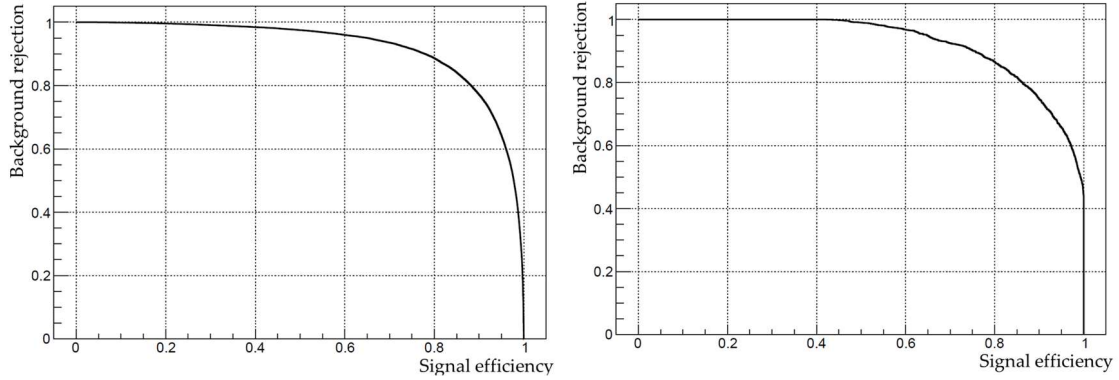

**Figure 6.** ROC curves for the NaF (left) and the aerogel (right) radiator

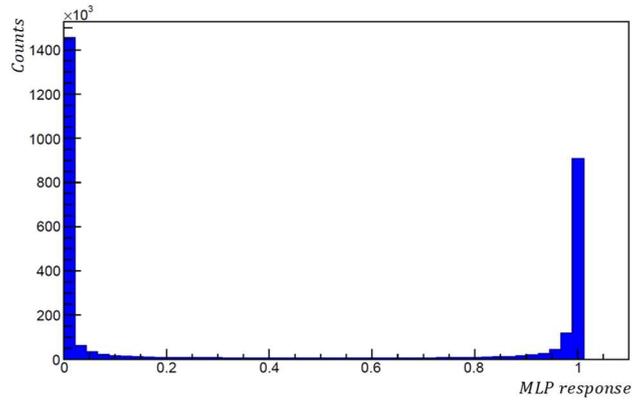

**Figure 7.** MLP score distribution for signal and background in the NaF radiator

After the training procedure is complete, the MLP-BFGS is applied on data collected by AMS-02 in the 2011-2021 period. The MLP cut has been implemented and the final mass distributions have been reconstructed. The results indicate that the MLP-BFGS classifier distinctly separates the signal and background peaks. Figure 8 illustrates the mass distribution for the events in the NaF and aerogel radiators, without selection and after the MLP cut.

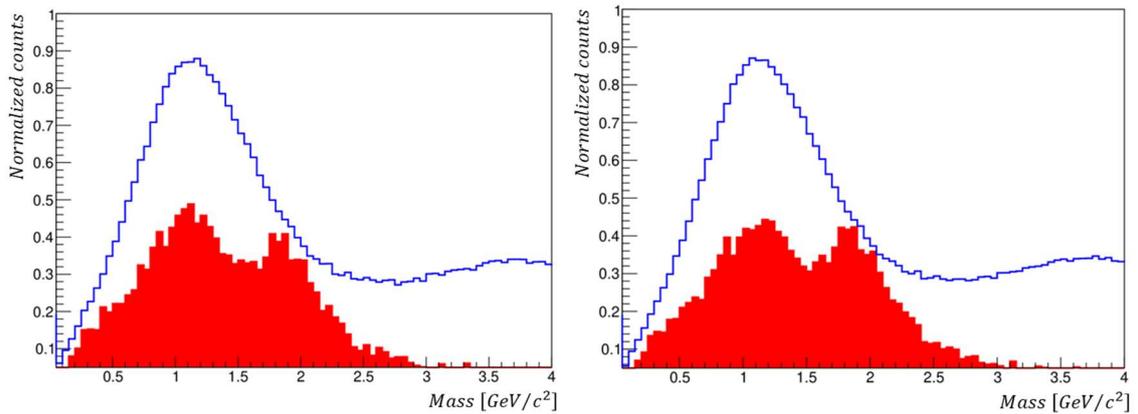

**Figure 8.** Normalized mass distributions of events in the NaF (left) and aerogel (right) radiators, without selection cuts (blue) and with a MLP cut (red)

The deuteron peak is clearly visible compared to the mass distribution before the cut, where the background contribution is such that a second peak is not noticeable at all.

## 5. Conclusion

Over the years, the measurement of charged cosmic ray fluxes has provided important information about the mechanisms of propagation of cosmic rays through the interstellar medium. The measurement of cosmic ray fluxes from low to extreme energies is linked to some of the most intriguing and still unsolved questions in physics, such as the nature of dark matter, the origin of matter-antimatter asymmetry, the existence of exotic new forms of matter, etc.

The identification of singly-charged isotopes is quite challenging due to the very large background contribution. Solving this problem in an efficient way is of great importance for the detection of antideuterons with energies below 2 GeV. It would be very powerful evidence for the existence of the supersymmetric neutralino, which is one of the main candidates for the dark matter particle.

In this study we show the results of MLP-BFGS training on simulation data and its testing on data collected by AMS-02 in the 2011-2021 period. The results indicate a clearly visible deuteron peak in the mass distribution, which is due to the successful rejection of a large number of background events. This study highlights the high potential of the MLP classifier as a tool for identification of signal and background, and to significantly improve particle identification by its mass.

## Acknowledgements

This publication is part of the project "Study of light antimatter production mechanisms in the Milky Way with the AMS detector" with project number № КП-06-Н58/11-23.11.2021, financed by the Bulgarian Scientific Research Fund (ФНИ). We are grateful for their support.

# References

[1] Planck Collaboration, P. A. R. Ade et al., *Planck 2015 results. XIII. Cosmological parameters*, Astron. Astrophys. 594 (2016) A13, arXiv:1502.01589

[2] R. H. Cyburt, B. D. Fields, K. A. Olive, T.-H. Yeh, *Big Bang Nucleosynthesis: 2015*, Rev. Mod. Phys. 88 (2016) 015004, arXiv:1505.01076

[3] *Anti-deuterons as a signature of supersymmetric dark matter*, Phys.Rev.D 62 (2000) 043003

[4] de Nolfo G.A., et al., *A measurement of cosmic ray deuterium from 0.5-2.9 GeV/nucleon*, AIP Conf. Proc., 528 (1), pp. 425-428, 10.1063/1.1324352, 2000

[5] Kim K., et al., *Cosmic ray 2H/1H ratio measured from BESS in 2000 during solar maximum*, Adv. Space Res., 51 (2), pp. 234-237, 2013

[6] Papini P., et al., *High-energy deuteron measurement with the CAPRICE98 experiment*, Astrophys. J., 615, pp. 259-274, 10.1086/424027, 2004

[7] Adriani O., et al., *Measurements of cosmic-ray hydrogen and helium isotopes with the PAMELA experiment*, Astrophysical Journal Volume 818, Issue 1, Article number 68, 2016

[8] The AMS collaboration, *The Alpha Magnetic Spectrometer (AMS) on the international space station: Part II - Results from the first seven years*, Phys. Rep. **894**, 1 (2021)

[9] P. Zuccon, *AMS-02 Track reconstruction and rigidity measurement*, 33rd International Cosmic Ray Conference, p. 1064 (2013)

[10] J. Burger, S. Gentile on behalf of the AMS-02 TRD Group, *The Performance of the AMS-02 TRD*, Proceedings of the 28th International Cosmic Ray Conference, 2003

[11] R. K. Hashmani, E. Akbas, M. B. Demirköz, *A comparison of deep learning models for proton background rejection with the AMS electromagnetic calorimeter*, Machine Learning: Science and Technology, Volume 5, Number 4, 2024

[12] A. Kounine, Z. Weng, W. Xu, C. Zhang, *Precision measurement of 0.5 GeV–3 TeV electrons and positrons using the AMS Electromagnetic Calorimeter*, Nuclear Instruments and Methods in Physics; Research Section A: Accelerators, Spectrometers, Detectors and Associated Equipment, Volume 869, p. 110-117, 2017

[13] A. Kounine, Z. Weng, W. Xu, C. Zhang, *Precision measurement of 0.5 GeV – 3 TeV electrons and positrons using the AMS electromagnetic calorimeter*, Nucl. Instrum. Methods Phys. Res. A 869 110-7, 2017

[14] D. Bourilkov, *Machine and Deep Learning Applications in Particle Physics*, International Journal of Modern Physics A, Vol. 34, No. 35, 1930019, 2019

[15] R. Pereira, on behalf of the AMS RICH collaboration, *The RICH detector of the AMS-02 experiment: status and physics prospects*, LIP/IST, arXiv:0801.3250v1 (astro-ph), 2018

[16] E. F. Bueno, F. Barão, M. Vecchi, *Machine learning approach to the background reduction in singly charged cosmic-ray isotope measurements with AMS-02*, Nuclear Instruments and Methods in

Physics Research, Section A: Accelerators, Spectrometers, Detectors and Associated Equipment, Volume 1056, art. №168644, 2023

[17] A. Hoecker, P. Speckmayer, J. Stelzer, J. Therhaag, E. von Toerne, H. Voss, *TMVA 4: Toolkit for Multivariate Data Analysis with ROOT – Users Guide*, CERN-OPEN-2007-007, TMVA version 4.0.1, 2018

[18] *"ROOT Data Analysis Framework"*, https://root.cern.ch/

[19] W.T. Eadie, D. Drijard, F. E. James, *Statistical Methods in Experimental Physics*, North-Holland, Amsterdam, 1971

[20] C.G. Broyden, *The Convergence of a Class of Double-rank Minimization Algorithms*, J. Inst. of Math. and App. 6, 76, 1970;

[21] R. Fletcher, *A New Approach to Variable Metric Algorithms*, Computer J. 13, 317, 1970;

[22] D. Goldfarb, *A Family of Variable Metric Updates Derived by Variational Means*, Math. Comp. 24, 23, 1970;

[23] D.F. Shannon, *Conditioning of Quasi-Newton Methods for Function Minimization*, Math. Comp. 24, 647, 1970

[24] I. Goodfellow, Y. Bengio, A. Courville, *Deep Learning, Chapter 6.5: Back-Propagation and Other Differentiation Algorithms*, MIT Press, pp. 200–220. ISBN 9780262035613, 2016

[25] J. Allison, et al., *Recent Developments in Geant4*, Vol. 835, pp. 186–225, http://dx.doi.org/10.1016/j.nima.2016.06.125